\documentclass[twocolumn,showpacs,preprintnumbers,amsmath,amssymb]{revtex4}

\usepackage{array}
\usepackage{booktabs}
\usepackage{tabu}
\usepackage{dcolumn}
\usepackage{amsmath}
\usepackage{amsfonts}
\usepackage{amssymb}
\usepackage{graphicx}
\usepackage{subfigure}
\usepackage{graphicx}
\usepackage{dcolumn}
\usepackage{bm}
\usepackage{xcolor}

\def\be{\begin{equation}}
  \def\ee{\end{equation}}
\def\bea{\begin{eqnarray}}
\def\eea{\end{eqnarray}}
\def\f{\frac}
\def\n{\nonumber}
\def\l{\label}
\def\p{\phi}
\def\o{\over}
\def\R{\rho}
\def\pa{\partial}
\def\om{\omega}
\def\na{\nabla}
\def\P{\Phi}

\begin{document}

\title{Brane inflation and Trans-Planckian censorship conjecture}

\author{Abolhassan Mohammadi}
\email{a.mohammadi@uok.ac.ir; abolhassanm@gmail.com}
\author{Tayeb Golanbari}
\email{t.golanbari@gmail.com}
\author{Jamil Enayati}
\email{j.enayati@garmian.edu.krd}

\affiliation{
Department of Physics, Faculty of Science, University of Kurdistan,  Sanandaj, Iran.\\
Physics Department, College of Education, University of Garmian, Iraq. \\
}
\date{\today}

\def\be{\begin{equation}}
  \def\ee{\end{equation}}
\def\bea{\begin{eqnarray}}
\def\eea{\end{eqnarray}}
\def\f{\frac}
\def\n{\nonumber}
\def\l{\label}
\def\p{\phi}
\def\o{\over}
\def\R{\rho}
\def\pa{\partial}
\def\om{\omega}
\def\na{\nabla}
\def\P{\Phi}

\begin{abstract}
The constraint of trans-Planckian censorship conjecture on the brane inflation model is considered. The conjectures put an upper bound on the main parameter including temperature, inflation time, potential, and the tensor-to-scalar ratio parameter $r$. It is determined that the resulting constraint could be stronger than what we have for the standard inflationary models. The constraint, in general, depends on the brane tension and it is concluded that the conjecture also confined the value of brane tension to have consistency for the model. Confining the brane tension turns into a determining value for the five-dimensional Planck mass. The case of slowly varying Hubble parameter gives more interesting results for the $\epsilon$ and $r$, which indicates that the value of the tensor-to-scalar perturbation is not required to be extremely small.
\end{abstract}
\maketitle

\section{Introduction}
Understanding the origin of the initial condition of the universe is one of the main purpose of the cosmology. Inflationary scenario not only solves the problems of the standard big-bang theory, it also predicts the quantum perturbations which are the main seeds for large scale structure of the universe \cite{Weinberg:2008zzc,Lyth:2009zz,Riotto:2002yw,Baumann:2009ds}. The scenario has received huge interest since its introduction \cite{starobinsky1980new,Guth:1980zm,albrecht1982cosmology,linde1982new,linde1983chaotic} and it has been generalized in many differential ways \cite{Barenboim:2007ii,Franche:2010yj,Fairbairn:2002yp,Mukohyama:2002cn,Spalinski:2007dv,Bessada:2009pe,maeda2013stability,
berera1995warm,maartens2000chaotic,Motohashi:2014ppa}. \\
The scenario of inflation has been extremely supported by the observational data \cite{Planck:2013jfk,Ade:2015lrj,Akrami:2019izv}, and many of different inflationary models could successfully pass the observational test \cite{Unnikrishnan:2012zu,Rezazadeh:2014fwa,Cespedes:2015jga,Nazavari:2016yaa,Amani:2018ueu,tirandari2017anisotropic,
Mohammadi:2018zkf,taylor2000perturbation,Sayar:2017pam,Akhtari:2017mxc,Sheikhahmadi:2019gzs,golanbari2014brane,Mohammadi:2018oku,
Mohammadi:2019dpu,Mohammadi:2019qeu,
Mohammadi:2020ftb,Mohammadi:2020ftb,Mohammadi:2020ake}. Besides, there are some other conjectures that it is expected that they should be satisfied by an inflationary model. The story is that the general theory of relativity is a low-energy effective field theory where the scale of energy is the Planck mass. Based on string theory, which is known as the best candidate for quantum gravity, every effective field theory should possess some features and meets some conjecture. One series of these conjectures for dividing the consistent effective field theory from the inconsistent ones is the swampland conjectures proposed by \cite{Obied:2018sgi,Garg:2018reu,Ooguri:2018wrx} which has been the topic of many research works \cite{Kehagias:2018uem,Das:2018rpg,Kinney:2018kew,Matsui:2018bsy,Lin:2018rnx,Dimopoulos:2018upl,Kinney:2018nny,
Geng:2019phi,Brahma:2018hrd,Brahma:2019iyy,Wang:2019eym,Odintsov:2020zkl,Odintsov:2018zai,Rasheed:2020syk,Brahma:2020cpy}. The swampland conjectures concern field distance and de Sitter vacuum. An effective field theory that has a de Sitter vacuum is not consistent with quantum gravity and stands in the swampland zone. The swampland criteria have been considered for different inflationary models, and it seems that the standard inflationary model with a canonical scalar field is in direct tension with these conjectures. However, modified models of inflation, e.g. k-essence model \cite{Lin:2019pmj}, and warm inflation \cite{Das:2019hto} have the chance to
successfully pass the test. \\
At the time of inflation, the universe is dominated by a scalar field, named inflaton, which produces a quasi-de Sitter expansion. Then, the universe undergoes an extreme expansion in a short time. As inflation expands the spacetime, the quantum fluctuations are also stretched out and their wavelengths grow and cross the Hubble horizon, while the Hubble horizon remains almost unchanged. At the time that the quantum fluctuations cross the horizon, they freeze and lose their quantum nature. Standing on the same logic as the last paragraph, the Trans-Planckian censorship conjecture (TCC) has been proposed by Bedroya and Vafa \cite{Bedroya:2019snp} stating that no quantum fluctuation with a wavelength shorter than the Planck length is allowed to cross the Hubble horizon, freezes and become classical. The conjecture could be formulated as
\begin{equation}\label{TCCconjecture}
  e^N l_p < H_e^{-1}
\end{equation}
where $l_p$ is the Planck length and $H_e^{-1}$ is the Hubble horizon at the end of inflation. \\
The TCC imposes a constraint on the model which includes spacetime expansion, like inflation, and it implies no limit on the cosmological evolution of standard big-bang cosmology where the fluctuations never cross the horizon. For the inflationary model, the TCC leads to some severe constraints \cite{Bedroya:2019tba,Brandenberger:2019eni} which will challenge many of the current models of inflation. In \cite{Bedroya:2019tba}, it is shown that the consequence of TCC on the energy scale of inflation is
\begin{equation*}
  V_e < 10^{10} \; {\rm GeV},
\end{equation*}
which results to an upper bound on the tensor-to-scalar parameter $r$ \cite{Bedroya:2019tba},
\begin{equation*}
  r < 10^{-30},
\end{equation*}
which anticipates an extremely small amplitude for the primordial gravitational waves. Note that the result has been obtained by taking an almost constant Hubble parameter during inflation and it is assumed that right after inflation we have a radiation dominant phase and reheating occurs very fast. For the k-essence model, a generalized version of TCC is proposed which involves the sound speed of the model \cite{Lin:2019pmj} in which for $c_s < 1$ the constraint gets even stronger. The TCC is studied in warm inflation as well \cite{Das:2019hto}, where it is determined that, in contrast to the canonical cold inflation, the warm inflation in a strong dissipative regime could properly pass all three. The conjecture has been the topic of some other researches such as \cite{Mizuno:2019bxy,Shi:2020ymp,Torabian:2019zms,Kamali:2019gzr,
Schmitz:2019uti,Brahma:2019unn,Brahma:2019vpl,Berera:2020dvn}. \\
Here, we are going to consider the constraint of TCC on bran inflation. The obtained bound on the energy scale and $r$ depends on the Friedmann equation of the model. The equation is modified in higher dimension models of cosmology, in which in RS brane-world and extra quadratic terms of energy density also appears in the Friedmann equation which dominates the linear term in high energy regime. Due to this, the TCC imposes a stronger constraint on the parameter of brane inflation than the standard inflation. \\

\section{TCC in brane inflation}
This conjecture could be stated in the following form
\begin{equation}\label{TCC}
  {a_e \over a_i} \; l_p \leq H_e^{-1}
\end{equation}
where the subscribes $e$ and $i$ stand for end and beginning of inflation and $l_p$ indicates the Planck length; $l_p^{-1} = m_p$. Following \cite{Bedroya:2019tba}, this condition could take a stronger form. Suppose that there is an expanding phase for the universe in the pre-inflationary phase. Then, it is reasonable to assume that there might be some modes with a physical wavelength equal to or shorter than the Planck length, $l_p$, between the time $t_p$ (Planck time)and $t_i$. This lead one to an stronger version of the above conjecture as
\begin{equation}\label{TCCstronger}
  {a_e \over a_p} \; l_p \leq H_e^{-1}
\end{equation}
Because of having an inverse relationship between the scale factor and the temperature in the radiation dominant phase, the above condition is expressed in terms of the temperature as well
\begin{equation}\label{TCCstrongerTemp}
  {T_p \over T_i} \; e^{N} \; l_p \leq H_e^{-1}
\end{equation}
in which $T_p$ is the temperature at the Planck time and $T_i$ is the temperature at the beginning of inflation. \\
Same as \cite{Bedroya:2019tba}, it is assumed that the pre-inflationary phase is a radiation dominant era where the scale factor behaves as $a(t) \propto t^{1/2}$. On the other hand, there is another condition related to the possibility of producing a causal mechanism for the observed structure of the universe. It states that the current comoving Hubble radius must be originated inside the comoving Hubble radius at the onset of inflation. In a mathematical language, it means
\begin{equation*}
  \left( a_0 H_0 \right)^{-1}  \leq \left( a_i H_i \right)^{-1}
\end{equation*}
which after some manipulation could be rewritten as
\begin{equation}\label{LSScondition}
  H_0^{-1} \; e^{-N} \; {T_0 \over T_e} \leq H_i^{-1} = H_e^{-1}
\end{equation}
the last equality on the right hand side of the equation is based on our assumption for this section that the Hubble parameter remains constant during inflation. $T_0$ and $T_e$ are the temperature at the present time and at the end of inflation. And $H_0$ is the present Hubble parameter. \\
Combining the two conditions, Eq.\eqref{TCCstrongerTemp} and \eqref{LSScondition}, results in
\begin{equation}\label{HT}
  {H_e \over H_0} \; {T_0 \over T_e} \leq {1 \over H_i l_p} \; {T_i \over T_p}.
\end{equation}
Just before and right after inflation, there is an inflationary phase. Then, the energy density in the Friedmann equation is a thermal bath of radiation, so
\begin{equation}\label{Friedmann}
  H^2 = {8 \pi \over 3m_p^2} \; \rho_r \; \left( 1 + {\rho_r \over 2\lambda} \right)
\end{equation}
where $\rho_r$ is the energy density of the thermal bath given by
\begin{equation}\label{rhoRadiation}
  \rho_r = {\pi ^2 \over 30} \; g_\star(T) \; T^4.
\end{equation}
We restrict the situation to the high energy regime where $\rho_r \gg \lambda$. In this case, the Friedmann equation is simplified to
\begin{equation}\label{FriedmannHE}
  H = \sqrt{4 \pi \over 3m_p^2 \lambda} \;  {\pi^2 \over 30} \; g_\star(T) \; T^4
\end{equation}
Evaluation $H_i$ and $H_e$ from the above equation and substituting the result in Eq.\eqref{HT}, there is
\begin{equation}\label{TempCondition}
  T_i^3 T_e^3 \leq {2.7 \times 10^3 \over 4\pi^5} \; {m_p^2 \lambda \over g_\star(T_i) g_\star(T_e)} \; {H_0 \over T_0}.
\end{equation}
Taking a matter dominant phase for the present time, the current Hubble parameter is read as
\begin{equation}\label{H0T}
  H_0^2 = {T_{eq} \over 3 m_p^2} \; T_0^3
\end{equation}
where $T_{eq}$ is the temperature at the time when energy density of matter and radiation are equal. Also, in the late time, the universe is in low energy regime, and the modified Friedmann equation \eqref{Friedmann} comes back to the standard form. Applying this equation on Eq.\eqref{TempCondition} leads to
\begin{equation}\label{TendCondition}
  T_e^6 \leq {2.7 \times 10^3 \over 4\pi^5 \sqrt{3}} \; {m_p \lambda \over g_\star(T_i) g_\star(T_e)} \; \sqrt{T_{eq} T_0}
\end{equation}
note that here it is assumed that $T_i = T_e$. Taking $g_\star(T_i) = g_\star(T_e) \simeq 10^2$, it turns to
\begin{equation}\label{TempConditionlambda}
  T_e^6 \leq 2.6 \times 10^4 \lambda \; {\rm GeV^2}
\end{equation}
which depends on the values of the brane tension; note that the brane tension $\lambda$ has the dimension ${\rm M^4}$, so the dimension of temperature is right. The brane tension has not been determined accurately but there are some estimations about it. To reproduce the nucleosynthesis as in standard cosmology $\lambda \geq 1 \; {\rm MeV^4}$ and also various astrophysical applications implies that $\lambda \geq 5 \times 10^8 \; {\rm GeV^4}$.\\
There is some understanding about the reheating temperature, which here is shown by $T_e$. After inflation, the universe is very cold and almost empty of any particle. The reheating phase is an explanation (and a possible scenario) for creating particles warming up the universe after inflation. However, there are some restrictions about the final temperature of the universe at the end of reheating. The reheating temperature should be high enough to recover the hot big bang nucleosynthesis, and on the other hand, it should be small enough to prevent the creation of unwanted particles. These conditions are combined as
$10^{-3} \; {GeV} \; < T_r=T_e \; < \; 10^{10} \; {GeV}$.
To satisfy the condition, the brane tension should stand in the range $10^{-22} \; {GeV^4} \; < \lambda \; < \; 10^{55} \; {GeV^4}$, which is very wide range. \\

The upper bound on temperature $T_e$ also implies an upper bound for the potential of the inflation as
\begin{equation}\label{STCCpot}
  V_e <  \Big( 2.465 \times 10^4 \; {\rm GeV^{4/3}} \Big)\; \lambda^{2/3}
\end{equation}
in which for $\lambda = 5 \times 10^{13} \; {\rm GeV^4}$ the potential should satisfy the upper bound $V_e < 3.3 \times 10^{13} \; {\rm GeV^4}$, meaning that the potential could have the same order as the brane tension or it should be lower. The important point is that it was assumed that the whole process of inflation occurs at high energy limit where $\lambda \ll V$. From Eq.\eqref{STCCpot} it is realized that as the brane tension gets bigger the High energy regime assumption are more likely to be violated, for example for $\lambda = 10^{15} \; {\rm GeV^4}$ the potential should satisfy the condition $V_e < 2.4 \times 10^{14} \; {\rm GeV^4}$ which clearly violated the high energy assumption. However, there is chance for preserving the high energy regime assumption for lower magnitude of the brane tension. For instance, by taking $\lambda = 10^{6} \; {\rm GeV^4}$ there is $V_e < 2.46 \times 10^8 \; {\rm GeV^4}$, which is consistent with the assumption. Therefore, it seems that the Eq.\eqref{STCCpot} applies a condition for the magnitude of brane tension as well. \\

How much expansion do we have in the inflationary phase? Substituting Eq.\eqref{STCCpot} in the TCC condition \eqref{TCC}, one finds a bound for the number of e-fold as
\begin{equation}\label{TCCefold}
  e^N < {2.85 \times 10^{33} \; {\rm GeV^{2/3}} \over \lambda^{1/6}}.
\end{equation}
Based on the wide studies of inflation, it is expected to have about $55-65$ number of e-fold expansion. Also Eq.\eqref{TCCefold} implies that lower brane tension leads to a bigger upper bound, in which for $\lambda = 10^{6} \; {GeV^4}$, we have $e^N < 2.8 \times 10^{32}$ meaning that $N < 74$ that clearly satisfies the aforementioned e-folding assumption.\\

The result for the temperature determines the values of the Hubble parameter $H_i$ which appears in the amplitude of the scalar perturbations as \cite{maartens2000chaotic,Wands:2000dp}
\begin{equation}\label{ps}
  \mathcal{P}_s = {3 \over 25\pi^2} \; \sqrt{4\pi \over 3 m_p^2 \lambda} \; {H_i^3 \over \epsilon},
\end{equation}
where $\epsilon$ is the first slow-roll parameter. According to the Planck data, the amplitude of the scalar perturbation is of the order of $\mathcal{P}_s \propto 10^{-9}$. To satisfy this observational constraint and at the same time hold the condition \eqref{TendCondition}, the slow-roll parameter should be about
\begin{equation}\label{epsilon}
 \epsilon \leq 1.277 \times 10^{-55}.
\end{equation}
The above constraint on the slow-roll parameter $\epsilon$, has a direct impact on the tensor-to-scalar parameter $r$ which is related to $\epsilon$ as \cite{Brax:2004xh,Langlois:2000ns,Huey:2001ae}
\begin{equation}\label{r}
  r \leq 2.043 \times 10^{-54}.
\end{equation}
which states that $r$ is extremely small. \\

The obtained result could be utilized to have some understanding about the start time of inflation. Suppose that after Planck time and before inflation the universe stands in radiation dominant phase. In this phase, the scale factor depends on time as $a(t) \propto t^{1/4}$, and since the scale factor is written as the inverse of the temperature, the beginning time of inflation is achieved as
\begin{equation}\label{initialtime}
  t_i = {T_p^4 \over T_i^4} \; t_p \geq {2.94 \times 10^{73} \; {\rm GeV^{8/3}} \over \lambda^{2/3}} \; t_p
\end{equation}
It means that if one wants to have inflation started at the time about $10^{-36} {\rm s}$, then the brane tension should take a huge value as $\lambda \sim 10^{98} \; {\rm GeV^4}$. This value of the brane tension clearly violate the discussed high energy regime assumption. Preserving the assumption restricts the brane tension to almost be of the order of $\lambda \leq 10^{10} \; {\rm GeV^4}$. By inserting this value in Eq.\eqref{initialtime}, the start point of inflation is obtained to be about $t_i > 10^{23} \; {\rm s}$, which is absolutely unacceptable. \\


\section{TCC for varying Hubble parameter}
In this section, we relax the assumption of having a constant Hubble parameter during inflation. The Hubble parameter is assumed to vary slowly as Eq.\eqref{varyingH} which appears in the power-law models \cite{Lin:2019pmj}
\begin{equation}\label{varyingH}
  H(N) = H_i \; e^{\alpha N}
\end{equation}
where $N=0$ is associated to the start point of inflation and $\alpha$ is taken as constant. The TCC condition \eqref{TCCstronger} remains unchange and it is extracted that
\begin{equation}\label{TCCefold}
  e^{(1+\alpha)N} \leq {T_i \over T_p} \; {m_p \over H_i}
\end{equation}
The condition regarding the causal mechanism of the universe structure is read as
\begin{equation}\label{LSSconditionH}
   {H_e \over H_0} \; {T_0 \over T_e}  \leq   e^{(1+\alpha) N}
\end{equation}
Utilizing Eq.\eqref{TCCefold}, one arrives at
\begin{equation}\label{HTH}
   {H_e \over H_0} \; {T_0 \over T_e} \leq {1 \over H_i l_p} \; {T_i \over T_p}.
\end{equation}
Same condition as we had in the previous case for constant $H$. Using the Friedmann equation \eqref{FriedmannHE}, the above condition is given by Eq.\eqref{TempCondition}. The temperature $T_i$ is the radiation temperature at the start time of inflation. After inflation, we assumed that the reheating occurs fast and the universe is warm up to the temperature $T_{reh}$ and we take this temperature as the temperature of the universe at the end of inflation, $T_e = T_{reh}$. In the previous section, it was assumed that the temperatures at the end and beginning of inflation are the same. Here we suppose the relation $T_i = \beta T_e$ for these two temperature where $\beta$ is a constant. For the current Hubble parameter, Eq.\eqref{H0T} is applied. Therefore, one arrives at
\begin{equation}\label{TempConditionH}
  T_e^6 \leq {2.7 \times 10^3 \over 4\pi^5 \sqrt{3}} \; {m_p \lambda \over g_\star(T_i) g_\star(T_e) \; \beta^3} \; \sqrt{T_{eq} T_0}.
\end{equation}
or
\begin{equation}\label{TempConditionlambdaH}
  T_e^6 \leq \Big( 2.6 \times 10^4 \; {\rm GeV^2} \Big) \; {\lambda \over \beta^3} .
\end{equation}
which depends on the brane tension and the constant $\beta$. In comparison to Eq.\eqref{TempConditionlambda}, one could get a bigger values of temperature when $\beta<1$ resulting in $T_i < T_e$ which means that the temperature after inflation (provided by the reheating phase) is bigger that the temperature just before inflation. If $\beta > 1$ (i.e. $T_i > T_e$), the temperature receives a smaller values.\\

The upper bound \eqref{TempConditionlambdaH} leads to the following condition for the brane inflation
\begin{equation}\label{STCCpot}
  V_e < \Big( 2.465 \times 10^4 \; {\rm GeV^{4/3}} \Big)\; {\lambda^{2/3} \over \beta^2}
\end{equation}
which leads to the initial potential
\begin{equation}\label{STCCpotInitial}
  V_i < \Big( 2.465 \times 10^4 \; {\rm GeV^{4/3}} \Big)\; \beta^2 \; \lambda^{2/3}
\end{equation}
From the Friedmann equation \eqref{Friedmann} and Eq.\eqref{varyingH}, one finds that $V(N) = V_i \; e^{\alpha N}$. Comparing it with above relation, it is realized that the parameter $\alpha$ and $\beta$ are related via $\beta^2 = e^{-\alpha N}$. \\
Although higher values of brane tension $\lambda$ leads to the bigger inflation potential, it could weaken the high energy assumption. The difference with the previous case is the constant $\beta$ which could play an essential role for preserving the condition $V \ll \lambda$.
Eq.\eqref{STCCpotInitial} is reflected in the Hubble parameter $H_i$, which appears in the amplitude of the scalar perturbations. To come to an agreement with observational data there should be $\mathcal{P}_s \propto 10^{-9}$, the first slow-roll parameter $\epsilon$ should satisfy the following condition
\begin{equation}\label{epsilonH}
  \epsilon \leq 1.277 \times 10^{-55} \; \beta^6
\end{equation}
and this condition is reflected in the parameter $r$ as
\begin{equation}\label{rH}
  r \leq 2.043 \times 10^{-54} \; \beta^6
\end{equation}
Again, the large value of the constant $\beta$ could assuage the strong condition that we have for the previous case. The constant $\beta$ could lead to some interesting results. For instance, taking $\lambda = 2 \times 10^{30} \; {\rm GeV^4}$ and $\beta = 5 \times 10^8$ indicates that the reheating temperature and the inflation energy scale are respectively about $T_e = 26.26 \; {\rm GeV}$ and $V_i = 9.78 \times 10^{41} \; {\rm GeV^4}$. The reheating temperature stands in the acceptable range and the magnitude of the potential perfectly satisfies the high energy assumption. On the other hand, for these values of $\beta$, the slow-roll parameter $\epsilon$ and tensor-to-scalar ratio are estimated as $\epsilon = 0.00199$ and $r = 0.0319$, which states that to satisfy the TCC, the parameter $r$ is not required to be extremely small.


\section{Conclusion}
The recently proposed Trans-Planckian censorship conjecture seems to impose a strong constraint on standard inflation. The conjecture has been considered in brane inflation where there is an extra infinite spatial dimensional leading to a modified Friedmann equation. The Friedmann evolution equation of the model shows that there is an extra quadratic term of the energy density which dominates the linear term in the high energy regime. \\
It was assumed that there is a radiation-dominated epoch in the pre-inflationary phase and after inflation, we have again another reheating epoch. The reheating phase was assumed to occur very fast. The conjecture was considered for two cases first by taking the Hubble parameter as a constant during inflation and in the second case, it was taken as a slowly varying function. The TCC forbids any mode with an initial wavelength smaller or equal to the Planck length to cross the Hubble horizon and be classical. Applying the condition led to an upper bound for the temperature $T_e$, which in general depends on the brane tension. The temperature $T_e$ is also recognized as the reheating temperature, and due to our understanding of the reheating temperature, we found an acceptable range for the brane tension $\lambda$. The parameter affects the value of the potential as well, and preserving the high energy condition restricts the obtained range of the brane tension. Comparing the results with the case of the standard inflation, there are stronger constraints on $\epsilon$ and $r$ as $r < 10^{-54}$. However, a problem was encountered regarding the beginning time of inflation. Selecting $\lambda = 10^{10} \; {\rm GeV^4}$ implies a beginning time about $t_i = 10^{23} \; {\rm s}$, which is unacceptable. Note that higher values of $\lambda$ indicate smaller $t_i$, but it violates the high energy condition assumption, and smaller $\lambda$ predicts bigger $t_i$. \\
Next, we consider the case of slowly varying Hubble parameter. The results for the case were more interesting in which for the obtained range of the brane tension, the high energy condition is preserved and at the same time having a small initial time, about $10^{-8} {\rm s}$. The interesting point is that the tensor-to-scalar ratio parameter is not required to be extremely high, and it could be of the order of $10^{-2}$.

\section*{acknowledgement}
The authors thank \textit{Prof. Robert Brandenberger} for his valuable comments and the time he spent reviewing the manuscript. \\
The work of A.M. has been supported financially by "Vice Chancellorship of Research and Technology, University of Kurdistan" under research Project No.99/11/19063. The work of T. G. has been supported financially by "Vice Chancellorship of Research and Technology, University of Kurdistan" under research Project No.99/11/19305.










\bibliography{TCCbraneRef}



\end{document}